# Ising Superconductivity in Transition Metal Dichalcogenides

Noah F.Q. Yuan, Benjamin T. Zhou, Wen-Yu He and K. T. Law

DEPARTMENT OF PHYSICS, HONG KONG UNIVERSITY OF SCIENCE AND TECHNOLOGY,

CLEAR WATER BAY, HONG KONG, CHINA

**Abstract**

In this work, we review the results of several recent works on the experimental and theoretical studies of monolayer superconducting transition metal dichalcogenides (TMD) such as superconducting $MoS_2$ and $NbSe_2$. We show how the strong Ising spin-orbit coupling (SOC), a special type of SOC which pins electron spins to out-of-plane directions, can affect the superconducting properties of the materials. Particularly, we discuss how the in-plane upper critical fields of the materials can be strongly enhanced by Ising SOC and how TMD materials can be used to engineer topological superconductors and nodal topological superconductors that support Majorana fermions.

**Introduction**

Graphene is a two-dimensional sheet of carbon atoms with a honeycomb lattice structure [1]. There has been immense interest in the study of graphene over the past decade due to its two-dimensional nature, novel mechanical, optical and electrical properties, as well as it being a platform for studying Dirac fermions [2]. More recently, another two-dimensional material, monolayers of transition metal dichalcogenides (TMDs) such as $MoS_2$, $WSe_2$, $NbSe_2$, have been successfully isolated [3, 4]. A monolayer of TMD is formed by a layer of transition metal atoms with a triangular lattice sandwiched by two identical layers of chalcogen atoms with a triangular lattice as shown in Fig. 1a). When viewed from the top, a monolayer TMD has a honeycomb lattice structure like graphene, but with a broken A-B sublattice symmetry (Fig. 1 b)) and it has a massive Dirac energy spectrum. Due to the energy gap and relatively high mobility, TMDs

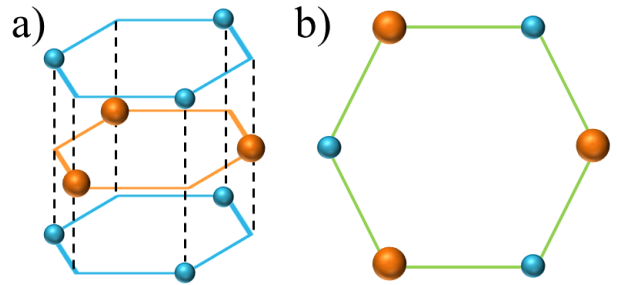

**Fig. 1:** a) Lattice structure of monolayer TMD, and b) its view from the top. The orange dots refer to the transition metal atoms, and the blue dots refer to the chalcogen atoms.

are considered as candidate materials for next generation transistors [5-9].

Importantly, unlike graphene, the atomic spin-orbit couplings (SOCs) in TMDs can be very strong due to the heavy transition metal atoms. Coupled with the fact that an in-plane mirror symmetry is broken in the lattice structure, the energy bands can be strongly split by SOCs [10-14].

Intuitively, the SOC experienced by a moving electron with momentum $\boldsymbol{k}$ is proportional to $\boldsymbol{k} \times \boldsymbol{E} \cdot \boldsymbol{\sigma}$ where $\boldsymbol{E}$ is the electric field experienced by the electron. Here, $\boldsymbol{\sigma}$ denotes the Pauli matrices. In a monolayer TMD, the in-plane mirror symmetry is broken and the electrons can experience in-plane electric fields. As a result, the SOC field has the form $S(\boldsymbol{k})\sigma_z$ and it pins electron spins to the out-of-plane directions where $S(\boldsymbol{k})$ is a function that depends on the lattice structure. We call this type of SOC the Ising SOC to distinguish it from the Rashba SOC, which has the form $k_x\sigma_y - k_y\sigma_x$ and pins electron spins to in-plane directions.

The strength of Ising SOCs in many TMD materials is very strong, in the order of hundreds of meV. In tungsten based materials such as WSe$_2$, the energy splitting on the top of the valence band can be as large as 450 meV. Due to time-reversal symmetry, the SOC has opposite signs for electrons with opposite momentum. As a result, electrons at opposite $K$ points experience opposite effective Zeeman fields. This is the so-called spin-valley locking effect. The importance of the spin-valley locking effect has been intensively studied in recent optical and transport experiments [15].

Interestingly, many TMD materials are superconducting. It is well-known that bulk NbSe$_2$, NbS$_2$, TaSe$_2$ and TaS$_2$ are intrinsic superconductors [16]. Even though MoS$_2$, MoSe$_2$, and WSe$_2$ are insulators, they can be superconducting when the bulk materials are intercalated by alkaline metals. However, bulk TMDs have rather different band structures from monolayer TMDs. The question is as follows: can monolayer TMDs be superconducting?

In 2012, it was demonstrated that a thin film (about 20 atomic-layers thick) of MoS$_2$ can be superconducting when all the conducting electrons are gated to the top atomic layer of the sample [17, 18]. The $T_c$ of the material can be as high as 10K at optimal gating. This discovery was significant because, unlike the bulk TMDs (with 2$H$ structure) that preserve inversion symmetry, monolayer TMDs break inversion symmetry and are non-centrosymmetric superconductors. Moreover, the SOCs in monolayer TMDs are strong, and non-trivial superconducting states may arise due to SOC.

To study the superconducting properties of gated MoS$_2$, the phase diagram of the material as a function of the interaction strengths of the electrons was worked out [19]. As expected, due to the strong SOC in the system, several interesting phases, including a so-called topological spin-singlet $p$-wave and a spin-triplet $s$-wave pairings phase were found [19, 20]. However, interactions beyond electron-phonon interactions are needed to realize these novel phases. The question is as follows: can strong Ising SOCs induce novel properties in a superconducting state even if the pairing is a conventional $s$-wave pairing? The answer is a definite yes.

After gated MoS$_2$ were found to be superconducting, two experimental groups independently discovered that the in-plane upper critical fields ($B_{c2}$) of the material were several times higher than the bulk materials and six times higher than the Pauli limits [21, 22]. In general, spin flip scatterings can enhance $B_{c2}$. However, these extremely high $B_{c2}$ cannot be explained by the Klemm-Luther-Beasley theory [23], which requires a spin-orbit scattering rate higher than the total scattering rate to explain the experimental data [21].

In fact, it was pointed out that the enhanced $B_{c2}$ is caused by the Ising SOC, which pins the electron spins to the out-of-plane directions such that in-plane magnetic fields cannot effectively polarize electron spins to in-plane directions [21, 22]. We call these superconductors with strong Ising SOCs, Ising superconductors.

Concurrent with developments in the gated MoS$_2$ case, monolayers of superconducting NbSe$_2$ were successfully fabricated with zero resistance $T_c$ of about 3K [24-27]. The superconductivity is undoubtedly two-dimensional as the samples are one-atomic-layer thick. It was found that the $B_{c2}$ is also about six times stronger than the Pauli limit [24]. The enhancement of $B_{c2}$ can be explained by taking into account all the Fermi pockets and the Ising SOC of the system [28].

It is important to note that the Rashba SOC in 2D superconductors can pin electron spins to in-plane directions so that an out-of-plane magnetic field cannot polarize the electrons. Therefore, Rashba superconductors are protected from paramagnetic effects of the magnetic field in the out-of-plane directions. However, out-of-plane fields always induce orbital effects so that the superconductivity can be destroyed by creating vortices. On the other hand, atomically thin Ising superconductors are protected from both orbital and paramagnetic effects of in-plane magnetic fields and result in extremely high in-plane upper critical fields.

Besides enhancing $B_{c2}$, are there any other novel properties of Ising superconductors? The answer here is also yes, and a few recent works address this question.

First of all, due to the Ising SOC, there can be mixing of spin-triplet Cooper pairs with spin-singlet Cooper pairs in superconductors, as expected by Gorkov and Rashba [29-31]. However, as we will demonstrate in the next section, the spin-triplet Cooper pairs are formed by electrons with the same spin polarization, which point to in-plane directions. Due to these equal-spin Cooper pairs, a triplet pairing gap can be induced on a wire through the proximity effect if the wire is placed on top of the Ising superconductor. This can result in a 1D topological superconducting wire with Majorana end states in the presence of an in-plane magnetic field [32]. Majorana fermions are quasiparticles that are their own anti-particles [33]. They satisfy non-Abelian statistics [34] and have potential applications in quantum computations [35]. This scheme of creating Majorana fermions is similar to the case of using Rashba wires on top of an $s$-wave superconductor [36-40]. The advantages of using Ising SOCs will be discussed in a later section.

Secondly, since $B_{c2}$ of the material can be much higher than the Pauli limit, one may ask whether the superconducting phase is changed under such strong magnetic fields. As we will explain below, for a monolayer NbSe$_2$, the in-plane magnetic field can drive the fully gapped $s$-wave pairing phase into a nodal topological phase through a topological phase transition when the applied in-plane magnetic field is larger than the Pauli limit but smaller than the $B_{c2}$. The in-plane magnetic field can create nodal points in the bulk energy spectrum and these nodal points are connected by Majorana flat bands [28], similar to Weyl points in Weyl semimetal being connected by surface Fermi arcs [41-43].

In the remaining sections, we will first explain the origin of the enhancement of $B_{c2}$. We will then discuss how Ising superconductors can be used to create Majorana fermions. Finally, we will briefly discuss how a monolayer of NbSe$_2$ can be driven to a nodal topological phase and describe the properties of this nodal topological phase.

## Ising Superconductivity of Gated Monolayer MoS$_2$

It has been reported that when the chemical potential of MoS$_2$ thin films are gated to the conduction band, the sample can become superconducting [17]. It is believed that the carriers are mostly concentrated at the top layer of the thin film, which makes the system an effective monolayer MoS$_2$. Importantly, it was shown that the in-plane $B_{c2}$ is strongly enhanced to several times the Pauli limit. The experimental results are shown in Fig. 3.

In order to understand the origin of the enhancement of in-plane $B_{c2}$, we first derive the Hamiltonian of the system and solve the self-consistent gap equation to obtain $B_{c2}$.

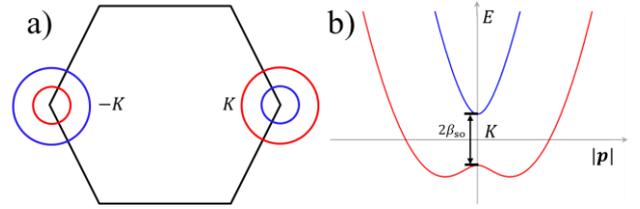

**Fig. 2:** a) The Brillouin zone and Fermi surfaces of a monolayer MoS2 near the $\pm K$ points. b) The energy spectrum of MoS$_2$. At K point, the band splitting is $2\beta_{so}$.

According to the first-principle calculations, the conduction band minima near the $\pm K$ points are dominated by the Mo $4d_{z^2}$ orbitals. Therefore, we denote the creation operators of $4d_{z^2}$ electrons as $c_s^\dagger$, where $s = \uparrow/\downarrow$ denotes spin. The gated monolayer MoS$_2$ respects the point group $C_{3v}$ symmetry. The normal Hamiltonian near the $\pm K$ points on the basis of $(c_{k\uparrow}, c_{k\downarrow})^\mathrm{T}$ can be written as [19]:

$$H_0(\boldsymbol{k} = \boldsymbol{p} + \epsilon \boldsymbol{K}) = \frac{|\boldsymbol{p}|^2}{2m} - \mu + \alpha_R \boldsymbol{g}(\boldsymbol{p}) \cdot \boldsymbol{\sigma} + \epsilon \beta_{so} \sigma_z$$

Here, $\epsilon = \pm$ is the valley index, $\boldsymbol{K}$ is the momentum of the $K$ point, $m$ is the effective mass of the electrons, $\mu$ is the chemical potential measured from the conduction band bottom when SOC is omitted. Due to gating, out-of-plan mirror symmetry is broken and Rashba SOC is present. The Rashba strength is denoted by $\alpha_R$ and the Rashba vector is $\boldsymbol{g}(\boldsymbol{p}) = p_y \boldsymbol{x} - p_x \boldsymbol{y}$. The Ising SOC strength is denoted by $\beta_{so}$ where $\beta_{so}$ is a constant. It is important to note that the Ising SOC depends on the valley index $\epsilon$ and it has opposite signs at opposite valleys. Therefore, electron spins at opposite valleys are polarized to opposite directions. The Fermi surfaces and the energy spectrum according to this normal Hamiltonian are depicted schematically in Fig. 2.

It is important to note that the Ising SOC is compatible with superconductivity. In general, the superconducting phase of the gated monolayer MoS2 can be described by the mean field BdG Hamiltonian

$$H_{\mathrm{BdG}}(\boldsymbol{k}) = \begin{pmatrix} H_0(\boldsymbol{k}) & \Delta_\Gamma(\boldsymbol{k}) \\ \Delta_\Gamma^\dagger(\boldsymbol{k}) & -H_0^*(-\boldsymbol{k}) \end{pmatrix}$$

in the Nambu basis $(c_{k\uparrow}, c_{k\downarrow}, c_{-k\uparrow}^\dagger, c_{-k\downarrow}^\dagger)^\mathrm{T}$. The mean field superconducting pairing matrix $\Delta_\Gamma(\boldsymbol{k})$ can be labeled by one irreducible representation $\Gamma$ of $C_{3v}$. As mentioned above, non-trivial pairing phases can appear but they require interactions beyond electron-phonon interactions [19].

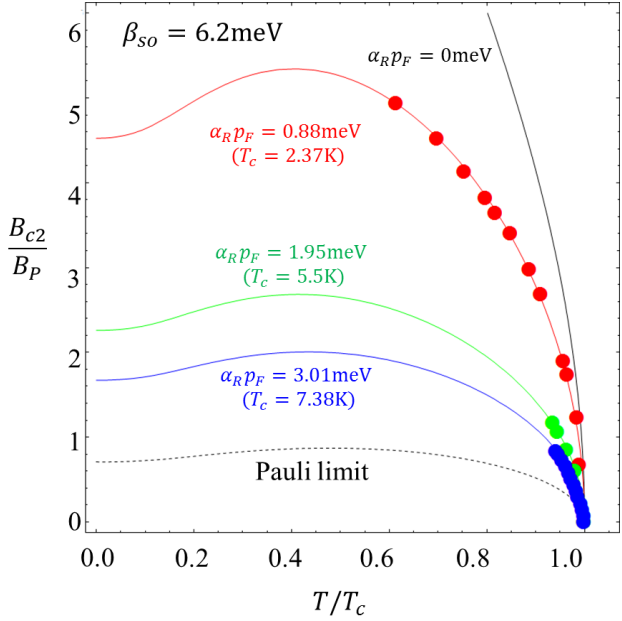

**Fig. 3**: Theoretical fitting of $B_{c2}/B_P$ versus $T/T_c$ relationship for experimental samples with $T_c$ = 2.37K (red), 5.5K (green) and 7.38K (blue), using a fixed Ising SOC ($\beta_{so}$ = 6.2meV) and varying Rashba SOC energy $\alpha_R p_F$, where $p_F$ is the Fermi momentum measured from $K$ points. The black dashed line depicts the results when both Ising SOC and Rashba SOC are zero and the conventional Pauli limit is recovered. The experimental data are extracted from Ref. [21].

By simply assuming a conventional $s$-wave pairing with $\Delta_\Gamma(\mathbf{k}) = \Delta_0 i\sigma_y$, we can solve the gap equation of $H_{\text{BdG}}$ from which we can then extract the $B_{c2}$ as a function of temperature. The results are depicted in Fig. 3.

The dots with different colors represent the $B_{c2}$ of samples with different $T_c$ extracted from Ref. [21]. The solid lines represent theoretical calculations. By assuming $\beta_{so}$ to be 6.2 meV, which causes an energy splitting of 12.4 meV of the energy bands at the $K$ points (Fig. 2b)), the experimental data can be well explained by also taking into account the effects of Rashba SOC. The Rashba SOC competes with the Ising SOC and lowers the $B_{c2}$. Moreover, the higher the $T_c$, the stronger the gating that is needed, resulting in a larger Rashba SOC. As shown in Fig. 3, the in-plane $B_{c2}$ can be as high as six times the Pauli limit.

## Ising Superconductors as a New Platform to Create Majorana Fermions

The search for Majorana fermions has been intense in condensed matter physics [44-47]. Majorana fermions are zero energy excitations in topological superconductors and they obey non-Abelian statistics. When two Majorana fermions exchange their positions, the quantum state of the system can be changed. This property of Majorana fermions makes them candidates for fault-tolerant quantum computation.

One of the most promising ways to realize topological superconductors is by placing a nanowire with Rashba SOC, such as an InSb wire, on top of a conventional $s$-wave superconductor and in the presence of a magnetic field [36-40]. The Rashba SOC plays the role of protecting and localizing the Majorana modes at the ends of the wire. Unfortunately, the Rashba SOC in InSb wires is rather small, in the order of 100 μeV [40]. Recently, we have proposed that Ising superconductors can be used to create topological superconductors by making use of the strong Ising SOC of the TMD materials.

The key observation is that Ising SOC induces spin-triplet Cooper pairs, even though the pairing potential is $s$-wave [32]. With the BdG Hamiltonian $H_{\text{BdG}}$ defined in the last section, one can explicitly work out the pairing correlation matrix $\hat{F}$ of the Ising superconductor defined as [29-32, 48]:

$$F_{\alpha\beta}(\mathbf{k}, E) = -i \int_0^\infty e^{i(E+i0^+)t} \langle \{c_{\mathbf{k},\alpha}(t), c_{-\mathbf{k},\beta}(0)\} \rangle dt$$

Here, $\alpha, \beta$ label the spin indices. For an arbitrary spin quantization axis which forms an angle $\theta$ with the $z$-axis, the pairing correlation matrix $\hat{F}_\theta(\mathbf{k}, E)$ becomes [32]:

$$\hat{F}_\theta(\mathbf{k}, E) = \begin{pmatrix} -d_z \sin\theta & \psi_s + d_z \cos\theta \\ -\psi_s + d_z \cos\theta & d_z \sin\theta \end{pmatrix}$$

where $\psi_s$ parametrizes the spin-singlet pairing correlation amplitude, and $d_z$ denotes the spin-triplet pairing correlation amplitude, which is generated by Ising SOC. In particular, when the spin quantization axis lies within the 2D plane of the Ising superconductor with $\theta = \frac{\pi}{2}$, all the triplet Cooper pairs are formed by equal-spin electron pairs with their spins pointing to in-plane directions as shown in Fig. 4.

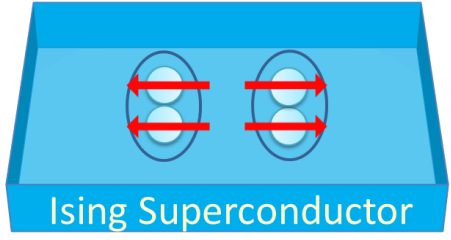

**Fig. 4:** The equal-spin triplet Cooper pairs generated by Ising SOC have spins pointing to in-plane directions.

As a result, when a paramagnetic wire is placed on top of the Ising superconductor and in the presence of an in-plane magnetic field, the electrons of the same spin near the Fermi energy of the wire can form pairs due to the equal spin-triplet pairing correlations induced by the Ising SOC. As a result, the wire can effectively become a 1-D $p$-wave superconductor [33] that supports Majorana end states.

The experimental setup of a nanowire (such as an InSb wire) placed on top of a superconducting TMD is shown in Fig. 5a). The spin of the wire at the Fermi energy is fully polarized by an in-plane external magnetic field. The spectral function of an infinitely long wire is shown in Fig. 5b). Evidently, the wire is fully gapped due to the superconducting proximity effect and the wire is equivalent to the Kitaev model of a 1D $p$-wave superconductor [33]. The energy spectrum of the superconducting TMD and a wire with finite length is shown in Fig. 5c). It is evident that for a wide range of chemical potentials, the system supports a zero energy fermionic mode (indicated by the red line). The wave function of a zero energy fermionic mode is shown in Fig. 5d). It is evident that this zero energy fermionic mode is non-local in the sense that the wave function is separated into two parts, and each part represents a Majorana fermion localized at one end of the wire.

One advantage of this setup is that the Ising SOC in the Ising superconductor is very strong and it is expected to induce strong SOC on the wire as well. Strong SOC can protect the Majorana modes from disorder effects. Secondly, due to the very high in-plane $B_{c2}$ of the Ising SOC, one may obtain a wider topological regime, as the width of the topological regime in Fig. 5c) is proportional to the Zeeman energy of the magnetic field. Another advantage of this setup is that the Majorana fermions can be created as long as the magnetic field is aligned in the in-plane directions [32]. Therefore, two wires on the Ising superconductor aligned in the T-junction geometry can be driven to the topologically non-trivial phase by an in-plane magnetic field simultaneously. On the contrary, in the Rashba wire case [36-39], an in-plane magnetic field is limited to aligning along the nanowire in order to create Majorana fermions. This property of Ising superconductors can be important for braiding Majorana fermions when T-junctions are needed [49].

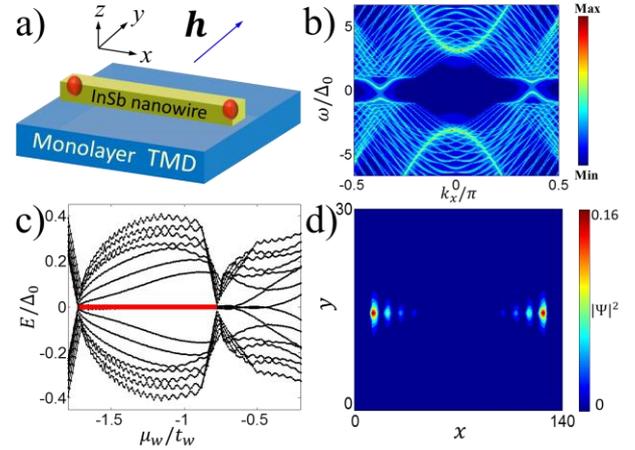

**Fig. 5:** a) A nanowire is placed on top of a superconducting TMD in the presence of in-plane external magnetic field $\boldsymbol{h}$. Majorana fermions (red dots) appear at the ends of the wire in the topological regime when all the electron spins in the nanowire are polarized along the magnetic field. b) The spectral function of the nanowire as a function of $k_x$. A superconducting gap is induced on the wire even though the wire is fully spin-polarized. c) The energy spectrum of the set-up in a) as a function of the chemical potential of the wire $\mu_w$ normalized by the band width of the wire. The topological regime is highlighted by the red line. d) The zero energy ground state of the system in the topological regime. Evidently, two Majorana fermions reside at the ends of the wire. Similar results can be found in Ref. [32].

Moreover, the nanowire on the Ising superconductor can be replaced by a chain of magnetic atoms and each atom can induce Yu-Shiba states [50,51]. These Yu-Shiba states can couple to each other and form an effective 1D topological superconductor [52-55].

**Nodal Topological Phase**

As discussed above, an important property of Ising superconductors is that the in-plane $B_{c2}$ can be strongly enhanced to several times higher than the Pauli limit. One more interesting question is as follows: how is the superconducting phase changed when the applied in-plane magnetic field is higher than the Pauli limit but below $B_{c2}$? Recently, it was shown that an in-plane magnetic field can drive the system from a fully gapped superconducting phase to a nodal topological phase [28].

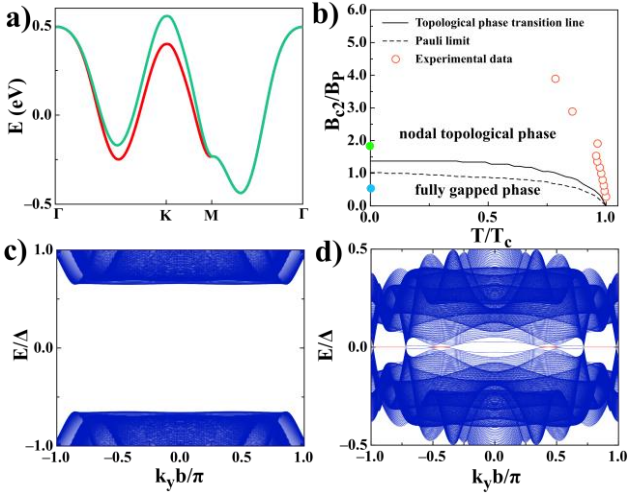

**Fig. 6:** a) The band structure for the normal state of monolayer NbSe$_2$ from a six band tight binding model described in Ref.[28]. b) The $B$-$T$ phase diagram of monolayer superconducting NbSe$_2$. It is evident that $B_{c2}$ is much higher than the Pauli limit. Here $B_{c2}$ is extracted from Ref. [24]. The area between the topological phase transition line and the $B_{c2}$ is the nodal topological phase regime. c) The energy spectrum of a strip of superconducting NbSe$_2$ with an armchair edge along the $y$-direction and an in-plane magnetic field $B = 0.5B_P$ [blue dot in b)]. d) Same as c) but $B = 1.8B_P$ [green dot in b)] such that the system is in the nodal topological regime. It is evident that the Majorana flat bands appear which connect the nodal points. Here $b = \sqrt{3}a$ and $a$ is the lattice constant.

More specifically, the phase diagram of monolayer NbSe$_2$ with the lattice structure shown in Fig. 1 was studied. The band structure near the Fermi energy is shown in Fig. 6(a). It is evident that the bands are strongly split due to Ising SOC. The $B_{c2}$ found in Ref. [24] is reproduced in Fig. 6(b). It is important to note that $B_{c2}$ is much higher than the Pauli limit, which is denoted by the dashed line.

To study the properties of the superconducting phase, the energy spectrum of a strip of NbSe$_2$ with armchair edges is calculated. The armchair edge is set along the $y$-direction. When the applied magnetic field is smaller than the Pauli limit, the energy spectrum is fully gapped as shown in Fig. 6(c). However, in the regime where the Zeeman energy caused by the in-plane field $B$ is larger than the pairing gap $\Delta_0$ (the green dot in Fig. 6b), where $B = 1.8B_P$), the bulk energy gap is closed. The bulk nodal points are manifested as gap closing points in the energy spectrum in Fig. 6d). Importantly, the gap closing points are connected by Majorana flat bands that are highlighted in red, similar to the Weyl points being connected by surface Fermi arcs in Weyl semimetals [40-42].

Moreover, the Majorana flat bands are associated with a large number of Majorana fermions residing on the edge of the superconductor. This leads us to our next question: how can we detect these Majorana fermions?

Indeed, when a normal lead is attached to the edge of the sample to form a tunneling junction with the superconductor, each Majorana mode can induce resonant Andreev reflections and cause zero bias conductance peaks in tunneling experiments [56-58]. This is very similar to the case of nodal $d$-wave superconductors, where the zero energy fermionic bound states on the [110] edge can be detected by tunneling experiments [59, 60].

More interestingly, due to the Majorana fermions on the edge, the Cooper pairs that tunnel from the lead to the superconductor must be spin-triplet Cooper pairs formed by electrons with the same spin polarization direction $\boldsymbol{n}$. The spin polarization direction is determined by the Majorana fermion wavefunction [61-63]. On the other hand, electrons with opposite spin-polarization $-\boldsymbol{n}$ in the lead are all reflected as electrons. Therefore, the tunneling current through the superconductor/normal lead junction are fully spin-polarized. In other words, the large number of Majorana modes on the superconductor edge act as a perfect spin filter that only allows electrons with spin polarization $\boldsymbol{n}$ to tunnel though the junction. We believe that this property of Majorana fermions may allow nodal topological superconductors such as NbSe$_2$ to have potential applications in superconducting spintronics [64, 65].

Furthermore, there are a large number of TMD materials which are not superconducting but have extremely large Ising SOC. One may also ask whether it is possible to make use of these materials to create Ising superconductors and nodal topological superconductors. Recently, it was shown theoretically that TMD materials can induce Ising SOC on a conventional superconducting thin film through the proximity effect and they can strongly enhance the in-plane $B_{c2}$ of the superconductor. Moreover, an in-plane magnetic field can drive the TMD/superconductor heterostructure into a nodal topological phase [66]. However, experimental efforts will be needed to verify this prediction.

## Conclusion

In this work, we have reviewed the results of several recent works on the study of superconducting TMDs. The enhancement of in-plane $B_{c2}$ in superconducting TMDs is explained. The possible realizations of Majorana fermions and superconducting nodal topological phases in TMD materials are also discussed.

**Acknowledgements:** KTL acknowledges the support of HKRGC and the Croucher Foundation through HKUST3/CRF/13G, 602813, 605512, 16303014 and a Croucher Innovation Grant.